\documentclass[11pt]{article}
\usepackage{natbib}
\usepackage[normalem]{ulem}
\usepackage{amssymb,amsmath,amsthm,latexsym,amsfonts,amscd,dsfont,enumerate}
\usepackage{a4wide,graphicx,tikz}

\usetikzlibrary{arrows,shapes,shadows,fit,backgrounds}
\tikzstyle{trader} = [circle, draw, top color=white, bottom color=blue!30, draw=blue!50!black!100, drop shadow, minimum height=4em]
\tikzstyle{bank} = [rectangle, draw, top color=white, bottom color=red!20, draw=red!50!black!100, drop shadow, rounded corners, minimum height=3em, text width=4em, text centered]
\tikzstyle{market} = [rectangle, draw, top color=white, bottom color=green!20, draw=green!50!black!100, drop shadow, rounded corners, minimum height=3em, text width=4em, text centered]
\tikzstyle{background} = [rectangle,fill=gray!10, inner sep=0.2cm, rounded corners=5mm]
\tikzstyle{line} = [draw, latex'-latex']
\tikzstyle{from} = [draw, latex'-]
\tikzstyle{to} = [draw, -latex']

\usepackage[]{hyperref}
 \hypersetup{
 colorlinks=true, 
 breaklinks=true, 
 urlcolor= blue, 
 linkcolor= blue, 
 bookmarksopen=true, 
 pdfauthor={Pallavicini, A., Brigo D.}
 }

\newtheorem{theorem}{Theorem}[section]
\newtheorem{remark}[theorem]{Remark}
\newtheorem{proposition}[theorem]{Proposition}

\newcommand{\Ex}[2]{\mathbb{E}_{#1}\!\left[\,#2\,\right]}

\newcommand{\ExF}[2]{\mathbb{E}\!\left[\,#2\,\left|{\cal F}_{#1}\right.\right]}
\newcommand{\ExG}[2]{\mathbb{E}\!\left[\,#2\,\left|{\cal G}_{#1}\right.\right]}
\newcommand{\ExFT}[3]{\mathbb{E}^{#2}\!\left[\,#3\,\left.\right|{\cal F}_{#1}\right]}

\newcommand{\QxCT}[3]{\mathbb{Q}^{#1}\!\left\{\,#2\,\left.\right|\,#3\,\right\}}

\newcommand{\ind}[1]{1_{\{#1\}}}

\newcommand{\lgd}{\mbox{L{\tiny GD}}}

\newcommand{\beq}[0]{\begin{equation}}
\newcommand{\eeq}[0]{\end{equation}}
\newcommand{\beqn}[0]{\begin{equation*}}
\newcommand{\eeqn}[0]{\end{equation*}}
\newcommand{\balign}[0]{\begin{aligned}}
\newcommand{\ealign}[0]{\end{aligned}}

\title{Impact of Multiple Curve Dynamics\\in Credit Valuation Adjustments under Collateralization\thanks{The opinions here expressed are solely those of the authors and do not represent in any way those of their employers. This paper is a refined version of the initial report \cite{BrigoPallavicini2013}.}}
\author{
Giacomo Bormetti\thanks{Scuola Normale Superiore Pisa and QUANTLab Pisa, {\tt  giacomo.bormetti@sns.it}}
\ \ \ 
Damiano Brigo\thanks{Imperial College London, {\tt damiano.brigo@imperial.ac.uk}}
\ \ \
Marco Francischello\thanks{Imperial College London, {\tt m.francischello14@imperial.ac.uk}}
\ \ \
Andrea Pallavicini\thanks{Imperial College London and Banca IMI Milan, {\tt a.pallavicini@imperial.ac.uk}}
}
\date{
\small First Version: September 5, 2014.  This version: \today
}

\begin{document}

\maketitle

\begin{abstract}
We present a detailed analysis of interest rate derivatives valuation under credit risk and collateral modeling. We show how the credit and collateral extended valuation framework in Pallavicini et al (2011), and the related collateralized valuation measure, can be helpful in defining the key market rates underlying the multiple interest rate curves that characterize current interest rate markets. A key point is that spot Libor rates are to be treated as market primitives rather than being defined by no-arbitrage relationships. We formulate a consistent realistic dynamics for the different rates emerging from our analysis and compare the resulting model performances to simpler models used in the industry.  We include the often neglected margin period of risk, showing how this  feature may increase the impact of different rates dynamics on valuation. We point out limitations of multiple curve models with deterministic basis considering valuation of particularly sensitive products such as basis swaps. We stress that a proper wrong way risk analysis for such products requires a model with a stochastic basis and we show numerical results confirming this fact. 
\end{abstract}

{\bf JEL classification code: G13. \\ \indent AMS classification codes: 60J75, 91B70}

\medskip


{\bf Keywords:} Yield Curve Dynamics, Multiple Curve Framework, HJM Framework, Interest Rate Derivatives, Basis Swaps, Counterparty Credit Risk, Liquidity Risk, Funding Costs, Collateral Modeling, Overnight Rates.

\newpage
{\small \tableofcontents}
\vfill
\newpage

\pagestyle{myheadings} \markboth{}{{\footnotesize Bormetti, Brigo, Francischello, Pallavicini. ~Interest-Rate Modeling in Collateralized Markets.}}

\section{Introduction}
\label{sec:introduction}

After the onset of the crisis in 2007, all market instruments are quoted by taking into account, more or less implicitly, credit and collateral related adjustments. As a consequence, when approaching modeling problems one has to carefully check standard theoretical assumptions which often ignore credit and liquidity issues. One has to go back to market processes and fundamental instruments by limiting oneself to value only derivative contracts that can be replicated by means of market instruments. Referring to market observables and processes is the only means we have to validate our theoretical assumptions, so as to drop them if in contrast with observations. This general recipe is what is guiding us in this paper, where we try to adapt interest rate models for valuation to the current landscape. 

A detailed analysis of the updated valuation problem one faces when including credit risk and collateral modeling (and further funding costs) has been presented in \cite{brigo2014nonlinear2,brigo2015nonlinear}. We refer to those papers and references therein for a detailed discussion. Here we specialize our updated valuation framework to consider the following key points: 
\begin{itemize}
\item[(i)] Focus on interest rate derivatives.
\item[(ii)] Understand how the updated valuation framework involving collateralized valuation measures can be helpful in defining the key market rates; these are the rates underlying the multiple interest rate curves that characterize current interest rate markets; in this setting, spot Libor rates are taken as primitives and are not defined by no-arbitrage relationships, while both Libor and overnight indexed swaps (OIS) forward rates are obtained by zeroing specific contract values. 
\item[(iii)] Formulate a consistent realistic dynamics for the rates emerging from the above analysis and compare valuation based on these models to valuation based on more standard market models. 
\item[(iv)] Show how the framework can be applied to valuation of particularly sensitive products such as basis swaps under credit risk, collateral and margin period of risk. By margin period of risk we mean the risk that the position changes value dramatically during the time between the relevant counterparty default and the actual liquidation rather than to instantaneous contagion (for the latter see for example \cite{BrigoCapponiPallavicini}); while the latter is usually called gap risk, in this paper we will use ``gap risk'' in the sense of margin period of risk above. 
\item[(v)] Illustrate numerically how gap risk may help appreciate the impact of different interest-rate dynamics on valuation.
\item[(vi)] Point out limitations in some current market practices such as explaining the multiple curves through deterministic fudge factors or shifts where the option embedded in the CVA calculation would be priced without any volatility. 
\end{itemize}

Overall, this paper presents a rigorous attempt to connect the updated credit and collateral inclusive valuation paradigms to multiple-curves interest rate theory, thus encompassing two of the key post-2007 developments in derivatives markets. 
 This paper is an extended and refined version of ideas originally appeared in \cite{BrigoPallavicini2013}.

\section{Valuation Equation with Credit and Collateral}
\label{sec:pricing}

Classical interest-rate models were formulated to satisfy no-arbitrage relationships by construction, which allowed one to price and hedge forward-rate agreements in terms of risk-free zero-coupon bonds

\subsection{Spot Libor Rates as Market Primitives}

Starting from summer 2007, with the spreading of the credit crunch, market quotes of forward rates and zero-coupon bonds began to violate usual no-arbitrage relationships. The main driver of such behavior was the liquidity crisis reducing the credit lines along with the fear of an imminent systemic break-down. As a result the impact of counterparty risk on market prices could not be considered negligible any more.

This is the first of many examples of relationships that broke down with the crisis. Assumptions and approximations stemming from valuation theory should be replaced by strategies implemented with market instruments. For instance, inclusion of credit valuation adjustments (CVA) for interest-rate instruments, such as those analyzed in \cite{BrigoPallavicini2007}, breaks the relationship between risk-free zero coupon bonds and Libor forward rates. Also, funding in domestic currency on different time horizons must include counterparty risk adjustments and liquidity issues, see \cite{Filipovic2012}, breaking again this relationship. We thus have, against the earlier standard theory,
\begin{equation}
\label{eq:liborbreaking}
L(T_0,T_1) \neq \frac{1}{T_1-T_0} \left( \frac{1}{P_{T_0}(T_1)} - 1 \right)
\,,\quad
F_t(T_0,T_1) \neq \frac{1}{T_1-T_0} \left( \frac{P_t(T_0)}{P_t(T_1)} - 1 \right)
\,,
\end{equation}%
where $P_t(T)$ is a zero coupon bond price at time $t$ for maturity $T$, $L$ is the Libor rate and $F$ is the related Libor forward rate. A direct consequence is the impossibility to describe all Libor rates in terms of a unique zero-coupon yield curve. Indeed, since 2009 and even earlier, we had  evidence that the money market for the Euro area was moving to a multi-curve setting, as widely described in the literature. See, for instance, \cite{Bianchetti2009}, \cite{Henrard2007,Henrard2009}, \cite{Mercurio2009}, \cite{PallaviciniTarenghi}, and \cite{Fujii2011}.

A further example of evolving assumptions is given by the use of collateralized contracts. The growing attention on counterparty credit risk is transforming Over The Counter (OTC) derivatives money markets. An increasing number of derivative contracts is cleared by CCPs, while most of the remaining contracts are traded under collateralization, regulated by Credit Support Annex (CSA). Both cleared and CSA deals require collateral posting, as default insurance, along with its remuneration. We cannot neglect such effects. We refer to the extensive work in \cite{BrigoPallavicini2014} for an analysis of the CCP case.

\subsection{The Valuation Framework}

In order to value a financial product (for example a derivative contract), we follow the approach of \cite{Perini2011}, and we take the risk-neutral expectation of all the discounted cash flows occurring after the trading position is entered. In particular, we include in the cash flow computation all coupons, dividends and premiums listed in the contract and occurring upon default (close-out procedure), along with the cash flows required by the collateral margining procedure. Here, we do not analyze the impact of additional costs due to the funding or hedging procedure. The impact of these costs in pricing interest-rate derivatives is discussed in \cite{BrigoPallavicini2013}.

\subsubsection{The Master Formula}

We refer to the two names involved in the financial contract and subject to default risk as investor (also called name ``I'', usually the bank) and counterparty (also called name ``C'', for example a corporate client, but also another bank). We denote by $\tau_I$,and $\tau_C$ respectively the default times of the investor and counterparty. We fix the portfolio time horizon $T>0$, and fix the risk-neutral valuation model $(\Omega,\mathcal{G},\mathbb{Q})$, with a filtration $(\mathcal{G}_t)_{t \in [0,T]}$ such that $\tau_C$, $\tau_I$ are $(\mathcal{G}_t)_{t \in [0,T]}$-stopping times. We denote by $\Ex{t}{\cdot}$ the conditional expectation under $\mathbb{Q}$ given $\mathcal{G}_t$, and by $\Ex{\tau_i}{\cdot}$ the conditional expectation under $\mathbb{Q}$ given the stopped filtration $\mathcal{G}_{\tau_i}$. We exclude the possibility of simultaneous defaults, and define the first default event between the two parties as the stopping time
\[
\tau := \tau_C \wedge \tau_I \,.
\]%
We will also consider the market sub-filtration $({\cal F}_t)_{t \ge 0}$ that one obtains implicitly by assuming a separable structure for the complete market filtration $({\cal G}_t)_{t \ge 0}$. ${\cal G}_t$ is then generated by the pure default-free market filtration ${\cal F}_t$ and by the filtration generated by all the relevant default times monitored up to $t$ (see for example \cite{BieleckiRutkowski2002}). 

Here, without an explicit derivation, we write the pricing formula for a derivative contract inclusive of collateralized credit and debit valuation adjustments (CVA/DVA) and margining costs. We address the readers to \cite{Perini2011,Perini2012} for a complete discussion. The pricing master formula is the following:
\begin{equation}
\label{eq:fundingpreview}
V_t := \ExG{t}{\Pi(t,T\wedge\tau) + \gamma(t,T\wedge\tau) + \ind{t<\tau<T} D(t,\tau) \theta_{\tau} } \,,
\end{equation}%
where
\begin{itemize}
\item $\Pi(t,T)$ is the sum of all discounted payoff terms in the interval $(t,T]$, without credit or debit risk and without collateral cash flows. In other terms, these are the financial instrument cash flows without additional risks.  
\item $\gamma(t,T)$ are the collateral margining costs discounted cash flows within the interval $(t,T]$,
\item $\theta_{\tau}$ is the on-default cash flow. It is primarily this term that originates the CVA and DVA terms, that may also embed collateral and contagion risk, see for example the case of credit derivatives described in \cite{BrigoCapponiPallavicini}.
\end{itemize}

When we say discounting above, we mean discounting at the risk-free rate $r$ associated with the risk-neutral measure. We therefore need to define the related stochastic discount factor $D(t,\tau)$. Notation $D(t,u;r)$, or simply $D(t,u)$, in general will denote the risk-neutral discount factor, given by the ratio 
\[
D(t,u;r) := D(t,u) := B_t/B_u
\;,\quad
d B_t = r_t B_t dt \,,
\]%
where $B$ is the bank-account numeraire, driven by the risk-free instantaneous interest rate $r$ and associated to the risk-neutral measure $\mathbb{Q}$. This rate $r$ is assumed to be $(\mathcal{F}_t)_{t \in [0,T]}$ adapted and it is the key variable in all pre-crisis term structure modeling, since all other interest rates and bonds could be defined as a function of $r$, see for example \cite{BrigoMercurio2006}.

Common market procedures, as we will see later on, may link the paths of the additional valuation terms $\gamma$, $\theta$ in Equation~\eqref{eq:fundingpreview} to the future paths of the price $V$ itself, specifying further Equation \eqref{eq:fundingpreview} as a recursive relationship. This feature is hidden in simplified approaches based on adding a spread to the discount curve to accommodate collateral and funding costs. A different approach is followed by \cite{Crepey2011}, who extends the usual risk-neutral evaluation framework to include many cash accounts accruing at different rates.

\subsubsection{Cash-Flows in Continuous Time}


The pricing master equation depends on the specific form of the contractual cash-flows and the margining procedure. In \cite{Perini2011,Perini2012} these terms are defined as a sum of individual flows over a discrete time-grid. Since we are going to adapt the master equation to interest-rate derivatives, where collateralization usually happens on a daily basis, we prefer to model cash flows as happening in a continuous time-grid, since this simplifies notation and calculations. Furthermore, we assume that collateral re-hypothecation is allowed, as done in practice (see \cite{BrigoCapponiPallaviciniPapatheodorou} for a discussion of re-hypothecation). Thus, we define the unknown terms in Equation~\eqref{eq:fundingpreview} as
\begin{eqnarray}
\label{eq:costspayouts}
\Pi(t,u) &:=& \int_t^u  D(t,v) \,d\pi_v, \ \ \  
\gamma(t,u) = \int_t^u  ( r_v - c_v ) C_v D(t,v) \,dv, \\  \nonumber
\end{eqnarray}%
where $\pi_t$ is the contractual coupon process, and the collateral rate is defined as
\[
c_t := c^+_t \ind{C_t>0} + c^-_t \ind{C_t<0}
\]%
with $c^\pm$ defined in the CSA contract. In general we may assume the processes $c^+,c^-$ to be adapted to the default-free filtration ${\cal F}_t$.

We can plug the above definitions into Equation \eqref{eq:fundingpreview}, and we are able to write the following continuous-time master equation for valuation.

\begin{proposition} {\bf (Master equation for valuation under credit risk and collateralization).}
Under the assumptions above, the valuation master equation in presence of collateral \eqref{eq:costspayouts} for the payout $\Pi$ in \eqref{eq:costspayouts} is given by
\begin{equation}
\label{eq:pricing}
V_t = \ExG{t}{\int_t^T D(t,u;r) \left( \ind{u<\tau} \,d\pi_u + \ind{\tau\in du} \theta_u + ( r_u - c_u ) C_u \,du \right) }
\end{equation}%
where the discount factors are defined as given by
\[
D(t,T;x) := \exp\left\{-\int_t^T x_udu\, \right\} \,.
\]%
\end{proposition}

Notice that the above Equation \eqref{eq:pricing} is not suited for explicit numerical evaluations, since the right-hand side is still depending on the derivative price via the indicators within the collateral rates, and possibly via the on-default term, leading to recursive non-linear features. We could resort to numerical solutions, as in \cite{Crepey2012c}, but, since our goal is valuing interest-rate derivatives, we prefer to further specialize the valuation equation for such deals.

In this first work we develop our analysis without considering a dependence between the default times if not through their spreads, or more precisely by assuming that the default times are ${\cal F}$-conditionally independent. Moreover, we assume that the collateral account and the on-default processes are ${\cal F}$-adapted. Thus, we can simplify the valuation equation given by \eqref{eq:pricing} by switching to the default-free market filtration. By following the filtration switching formula in \cite{BCJR2008}, we introduce for any ${\cal G}_t$-adapted process $X_t$ a unique ${\cal F}_t$-adapted process ${\widetilde X}_t$, defined such that
\[
\ind{\tau>t} X_t = \ind{\tau>t} {\widetilde X}_t .
\]%
Hence, we can write the pre-default price process as given by $\ind{\tau>t} {\widetilde V}_t = V_t$ where the right hand side is given in Equation~\eqref{eq:pricing} and where $\tilde{V}_t$ is ${\cal F}_t$ adapted.

\subsection{Close-Out and Margining Procedures}

We continue this section by specifying the close-out procedure occurring at default time, so that we can define the on-default cash flow $\theta_u$ appearing in Equation~\eqref{eq:pricing}. We consider two possibilities: (i) at default time the close-out procedure is completed without delay, or (ii) the procedure takes $\delta$ days to be completed.

\subsubsection{The Close-Out Procedure without Delay}

The derivative pricing Equation \eqref{eq:pricing} requires a specification of the close-out procedure occurring at default time. A first possibility is given by a close-out procedure that at default time is completed without delay. In the following section we relax this assumption.

The analysis of the ISDA documentation describing the close-out procedure in presence of collateralization is described in the approximation of no delay in \cite{BrigoCapponiPallaviciniPapatheodorou}. On default event $\tau$ the surviving party computes the close-out value $\varepsilon_\tau$, which represents the value of the remaining cash flows. If the surviving party is a net debtor, then she must pay the whole close-out value $\varepsilon_\tau$ to the defaulting party. On the other hand, if the surviving party is a net creditor, then she is able to recover only a fraction of her credits. The collateral account is used to reduce the exposure, but if it is not enough an unsecured claim is needed to get back the remaining part. Here, we do not derive the analysis of \cite{BrigoCapponiPallaviciniPapatheodorou}, but we simply quote their result to write the on-default term $\theta_\tau$ as
\begin{equation}
\label{eq:theta}
\theta_\tau(C,\varepsilon) := \varepsilon_\tau - \ind{\tau_C<\tau_I} \lgd_C (\varepsilon_\tau-C_\tau)^+ - \ind{\tau_I<\tau_C} \lgd_I (\varepsilon_\tau-C_\tau)^-
\end{equation}%
where $\varepsilon$ is the exposure computed by the surviving party (close-out amount), $\lgd\in[0,1]$ is the loss given default, and $(x)^+$ indicates the positive part of $x$ while $(x)^- := -(-x)^+$.

The close-out amount is not tightly defined by ISDA doscumentation. Usually a risk-free or a pre-default specification is adopted. See \cite{BrigoMoriniPallavicini2012} for a discussion. Here, we choose a risk-free version aware of contract collateralization, and we define
\begin{equation}
\label{eq:close-out}
\varepsilon_t := \ExG{t}{\int_t^T D(t,u;r) \,d\pi_u + ( r_u - c_u ) C_u \,du} \,.
\end{equation}%
We complete this list of assumptions with a description of the margining procedure. For ease of discussion we adopt the choice of \cite{BrigoPallavicini2013}, and we write
\begin{equation}
\label{eq:collateral}
C_t := \alpha_t \varepsilon_t
\end{equation}%
where $\alpha_t\in[0,1]$ is ${\cal F}_t$ adapted. Notice that the close-out and the collateral processes depend on each other. We can solve the system of equations and we obtain
\[
\varepsilon_t = \ExG{t}{\int_t^T D(t,u;(1-\alpha)r+\alpha c)) \,d\pi_u} \,.
\]%
Notice that interest-rate derivatives have ${\cal F}$-adapted cash flows, so that in this paper we can also write
\begin{equation}
\label{eq:close-out-explicit}
\varepsilon_t = \ExF{t}{\int_t^T D(t,u;(1-\alpha)r+\alpha c)) \,d\pi_u}
\end{equation}%

\subsubsection{Switching to Market Filtration}

Then, if we can disregard contagion risk, as discussed in \cite{BrigoCapponiPallaviciniPapatheodorou}, we have that the value of the contract is continuous at default time, namely
\[
\tilde{V}_{\tau-} = \tilde{V}_\tau
\]%
and we have the possibility to simplify Equation \eqref{eq:pricing} by switching from the filtration $\cal G$ to the market filtration $\cal F$.

In general for any ${\cal F}_u$-adapted process we can write for any time $t$ and $u$ such that $t\leq u$
\[
\ExG{t}{ \ind{\tau\in du} \ind{\tau_C<\tau_I} \phi_u } = \ind{\tau>t} du\,\ExF{t}{\lambda^C_u D(t,u;\lambda) \phi_u }
\]%
and
\[
\ExG{t}{ \ind{\tau\in du} \ind{\tau_I<\tau_C} \phi_u } = \ind{\tau>t} du\,\ExF{t}{\lambda^I_u D(t,u;\lambda) \phi_u } \,.
\]%
Furthermore, we introduce the pre-default intensity $\lambda_t^I$ of the investor and the pre-default intensity $\lambda_t^C$ of the counterparty as
\begin{eqnarray}
\label{eq:lambda}
\ind{\tau_I>t} \lambda_t^I \,dt &:=& \QxCT{}{\tau_I\in dt}{\tau_I>t,{\cal F}_t} \\\nonumber
\ind{\tau_C>t} \lambda_t^C \,dt &:=& \QxCT{}{\tau_C\in dt}{\tau_C>t,{\cal F}_t}
\end{eqnarray}%
along with their sum $\lambda_t$
\begin{equation}
\label{eq:lambda_sum}
\ind{\tau>t} \lambda_t := \ind{\tau>t} \lambda_t^I + \ind{\tau>t} \lambda_t^C
\end{equation}%
and we omit the tilde over the intensity symbols to lighten the notation. We can now switch to the default-free market filtration $\cal F$ by means of the following proposition.

\begin{proposition}{\bf (Master equation under $\cal F$-conditionally independent default times without gap risk).}
\label{prop:masterF}
If we assume $\cal F$-conditionally independent default times and a $\cal F$-adapted payout, we can specialize the valuation Equation \eqref{eq:pricing} when the close-out procedure given by Equation \eqref{eq:close-out} is completed without delay at default time (no gap risk). We get
\[
V_t = \ind{\tau>t}  {\widetilde V}_t
\]%
where the pre-default value of the derivative contract is given by
\begin{eqnarray}
\label{eq:masterF}
\widetilde{V_t}
& = & \varepsilon_t \\\nonumber
& - & \ExF{t}{ \int_t^T D(t,u;r+\lambda) \lambda_u^C(1-\alpha_u) \lgd_C (\varepsilon_u)^+ \,du} \\\nonumber
& - & \ExF{t}{ \int_t^T D(t,u;r+\lambda) \lambda_u^I(1-\alpha_u) \lgd_I (\varepsilon_u)^- \,du} \,.
\end{eqnarray}
\end{proposition}

\begin{remark}{\bf (The perfect collateralization case)}
We can further specialize the above equation to a relevant case: when the margining procedure is able to track the exposure in continuous time, namely when
\[
\alpha_t = 1 \,.
\]%
This case is an approximation for interest-rate liquid instruments widely adopted in the literature. See, for instance, \cite{BrigoPallavicini2013}. In this case from Equations~\eqref{eq:collateral}, \eqref{eq:close-out-explicit}, and~\eqref{eq:masterF} we get
\begin{equation}
\label{eq:perfect}
\widetilde{V_t} = \varepsilon_t = C_t = \ExF{t}{\int_t^T D(t,u;c) \,d\pi_u} \,.
\end{equation}
\end{remark}

\subsubsection{The Close-Out Procedure with Delay: Gap Risk}

Although the margining prcedure reduces the counterparty risk, there are still some risks that can arise on default due to some mismatch between the margin account and the contract valuation. This problem is discussed in detail in \cite{BrigoPallavicini2014} for both bilateral and centrally cleared contracts. These risks contribute to what is called ``gap risk''. We give examples of some of them:   
\begin{itemize}
\item we can have a risk due to mismatch between the collateral account and the value of the contract just before the default, in the sense that the margining procedure is not perfect and do not match exactly the value of the contract;
\item we can have a risk because the contract value jumps at default, so diminishing the impact of our collateral on default risk;
\item we can have a risk due to delays in the default procedure. In fact usually there is a delay of $\delta$ days, called ``cure period'', between the default event $\tau$ and the close-out cash-flows at time $\tau+\delta$. Thus, the change in value of the contract over this period is not accounted for in the collateral account.
\end{itemize} 

In practice to mitigate such risks bilateral and centrally cleared contracts implements margining procedures which includes many collateral accounts with specific margining procedures. We refer to \cite{BrigoPallavicini2014} for details.

We can distinguish three different collateral accounts: the margin account $M_t$, that plays the same role of the account $C_t$ from the previous sections, and two initial margin accounts $N^I_t$ and $N^C_t$, representing the amount of money that the investor and the counterparty respectively should post into segregated accounts to cover gap risks. Remembering that we embody the investor's perspective we have that $N^I_t\leq 0$, since we put money in the account that will be used by the counterparty in case of our default, and $N^C_t\geq 0$ since we will benefit from this money in case the counterparty defaults. We further suppose that these two accounts are not subject to default because are kept segregated so that the amount of money that forms the initial margin accounts cannot be rehypothecated. Notice that in this case, differently from the variation margin account that can be viewed as a netting of two accounts, the two accounts are kept distinct.

A detailed description of the close-out procedure occurring in presence of a cure period, when the margining procedure includes of variation and initial margin accounts, is presented in \cite{BrigoPallavicini2014}. Here, we write the final expression without an explicit derivation.
\begin{eqnarray}
\label{eq:close-outgap}
\theta_\tau(M,N,\varepsilon)
& := & \ExG{\tau}{\varepsilon_{\tau+\delta}(\tau,T)} \\\nonumber
&  - & \ExG{\tau}{\ind{\tau_C<\tau_I+\delta}\lgd_C(\varepsilon_{\tau+\delta}(\tau,T)-M_\tau-N^C_\tau)^+} \\\nonumber
&  - & \ExG{\tau}{\ind{\tau_I<\tau_C+\delta}\lgd_I(\varepsilon_{\tau+\delta}(\tau,T)-M_\tau-N^I_\tau)^-}
\end{eqnarray}%
where we change the definition of close-out amount to deal with coupon paid during the cure period, namely
\begin{equation}
\label{eq:close-out-gap}
\varepsilon_s(t,T) = \ExG{s}{\int_t^T D(t,u;r) \,d\pi_u + ( r_u - c_u ) C_u \,du} \,.
\end{equation}%

Then, as in the previous section we make an assumption for the margining procedure. For ease of discussion we adopt the simple choice
\begin{equation}
\label{eq:collateral-gap}
M_t := \alpha_t \varepsilon_t
\;,\quad
N^I_t := N^C_t := 0
\end{equation}%
where $\alpha_t\in[0,1]$ is ${\cal F}_t$ adapted. A concrete definition of the initial margin accounts can be found in \cite{BrigoPallavicini2014}.

Then, by switching to the default-free market filtration $\cal F$, we have the following proposition.

\begin{proposition}{\bf (Master equation under $\cal F$-conditionally independent default times with gap risk}
\label{prop:masterFgap}
If we assume $\cal F$-conditionally independent default times and a $\cal F$-adapted payout, we can specialize the valuation Equation \eqref{eq:pricing} when the close-out procedure given by Equation \eqref{eq:close-out-gap} is completed with a delay of $\delta$ days at default time (gap risk). We get
\[
V_t = \ind{\tau>t}  {\widetilde V}_t
\]%
where the pre-default value of the derivative contract is given by
\begin{eqnarray}
\label{eq:masterFgap}
\widetilde{V_t}
& = & \varepsilon_t(t,T) \\\nonumber
& - & \ExF{t}{\int_t^T D(t,u;r+\lambda) \lambda_u^{\delta,C} \lgd_C(\varepsilon_{u+\delta}(u,T)-\alpha_u\varepsilon_u(u,T))^+ \,du} \\\nonumber
& - & \ExF{t}{\int_t^T D(t,u;r+\lambda) \lambda_u^{\delta,I} \lgd_I(\varepsilon_{u+\delta}(u,T)-\alpha_u\varepsilon_u(u,T))^- \,du} \,.
\end{eqnarray}
where we indicated
\[
\lambda^{\delta,C}_u=\lambda^C_u+\lambda^I_t(1-D(t,t+\delta;\lambda^C)),
\]
\[
\lambda^{\delta,I}_u=\lambda^I_u+\lambda^C_t(1-D(t,t+\delta;\lambda^I)).
\]

\end{proposition}

\section{Valuing Collateralized Interest-Rate Derivatives}
\label{sec:ir}

As we mentioned in the introduction, we will base our analysis on real market processes. All liquid market quotes on the money market (MM) correspond to instruments with daily collateralization at overnight rate ($e_t$), both for the investor and the counterparty, namely
\[
c_t \doteq e_t \,.
\]%
Notice that the collateral accrual rate is symmetric, so that we no longer have a dependency of the accrual rates on the collateral price, as opposed to the general master equation case. Moreover we further assume
\[
r_t \doteq e_t \,.
\]%
This makes sense because being $e_t$ an overnight rate, it embeds a low counterparty risk and can be considered a good proxy for the risk-free rate $r_t$. In principle one could instead use the reference rate for secured transactions (e.g. Eurepo in the Euro area) but there are fewer liquid products with long maturity and hence it presents some difficulties in the calibration to market data. Here, we do not consider funding costs. We refer to \cite{BrigoPallavicini2013} for details.

We will describe some of these MM instruments, such as overnight indexed swaps (OIS) and interest-rate swaps (IRS), along with their underlying market rates, in the following sections. At the moment of writing this paper an important part of the market is moving from OTC contracts regulated by a (standardized) bilateral CSA to a market cleared by CCPs. We refer to the extensive treatment in \cite{BrigoPallavicini2014} for details. Here, we assume, as in most of the literature, that the collateralization procedure operates in continuous time for liquid MM instruments being able to remove all credit risk, and, moreover, thet gap risk is not present, so that we can use the perfect collateralization approximation of Equation~\eqref{eq:perfect} to price these derivatives. See \cite{Henrard2014} for details. See \cite{BrigoCapponiPallaviciniPapatheodorou} for a discussion on the impact of discrete-time collateralization on interest-rate derivatives.

\subsection{Overnight Rates and OIS}
\label{sec:ois}

The MM usually quotes the prices of collateralized instruments. In particular, a daily collateralization procedure is assumed, so that CSA contracts require that the collateral account is remunerated at the overnight rate ($e_t$). In particular, the MM usually quotes the price of OIS. Such contracts exchange a fix-payment leg with a floating leg paying the same overnight rate used for their collateralization, compounded daily. Since we are going to price OIS under the assumption of perfect collateralization, namely we are assuming that daily collateralization may be viewed as done on a continuous basis, we approximate also daily compounding in OIS floating leg with continuous compounding, which is reasonable when there is no gap risk. Hence the discounted payoff of an OIS with tenor $x$ and maturity $T=nx$ is given by
\[
\sum_{i=1}^n D(t,T-(n-i)x;e) \left( 1 + x K - \exp\left\{ \int_{T-(n-i-1)x}^{T-(n-i)x}  e_u \,du \right\} \right)
\]%
In particular in the one-period case we have:
\[
D(t,T;e) \left(1 + x K - \exp\left\{\int_{T-x}^T  e_u \,du \right\}\right)
\]%
where $K$ is the fixed rate payed by the OIS. Furthermore, we can introduce the (par) fix rates $K \doteq E_t(T,x;e)$ that make the one-period OIS contract fair, namely priced $0$ at time $t$. They are implicitly defined as
\[
{\widetilde V}^{\rm OIS}_t(K) := \ExF{t}{\left( 1 + x K - \exp\left\{\int_{T-x}^T  e_u \,du \right\} \right) D(t,T;e)}
\]%
with
\[
{\widetilde V}^{\rm OIS}_t( E_t(T,x;e) ) = 0
\]%
leading to
\begin{equation}
\label{eq:EtT}
E_t(T,x;e) := \frac{1}{x} \left( \frac{P_t(T-x;e)}{P_t(T;e)} - 1 \right)
\end{equation}%
where we define collateralized zero-coupon bonds\footnote{Notice that we are only defining a price process for hypothetical collateralized zero-coupon bond. We are not assuming that collateralized bonds are assets traded on the market.} as
\begin{equation}
\label{eq:zcbond}
P_t(T;e) := \ExFT{t}{}{D(t,T;e)} \,.
\end{equation}%

One-period OIS rates $E_t(T,x;e)$, along with multi-period ones, are actively traded on the market. Notice that we can bootstrap collateralized zero-coupon bond prices from OIS quotes.

\subsection{Libor Rates, IRS and Basis Swaps}
\label{sec:irs}

Libor rates ($L_t(T)$) used to be linked to the term structure of default-free interbank interest rates in a fundamental way. In the classical term structure theory, Libor rates would satisfy fundamental no-arbitrage conditions with respect to zero-coupon bonds that we no longer consider to hold, as we pointed out earlier in Equation~\ref{eq:liborbreaking}. We now deal with a new definition of forward Libor rates that may take into account collateralization. Libor rates are still the indices used as reference rate for many collateralized interest-rate derivatives (IRS, basis swaps, \ldots). IRS contracts swap a fix-payment leg with a floating leg paying simply compounded Libor rates. IRS contracts are collateralized at overnight rate $e_t$. 
Thus, an IRS discounted payoff with maturity $T=nx$ and tenor $x$ is given by
\[
\sum_{i=1}^n x D(t,T-(n-i)x;e) \left( K -  L_{T-(n-i-1)x}(T-(n-i)x) \right) \,.
\]%
In particular in the one-period case we have:
\[
D(t,T;e) x (K -  L_{T-x}(T))
\]%
where $K$ is the fix rate payed by the IRS. Furthermore, we can introduce the (par) fix rates $K=F_t(T,x;e)$ that make the one-period IRS contract fair, namely priced $0$ at time $t$. They are implicitly defined via
\[
{\widetilde V}^{\rm IRS}_t(K) := \ExF{t}{\left( x K - x L_{T-x}(T) \right) D(t,T;e)}
\]%
with
\[
{\widetilde V}^{\rm IRS}_t(F_t(T,x;e)) = 0
\]%
leading to the following definition of collateralized forward Libor rate
\[
F_t(T,x;e) := \frac{\ExF{t}{L_{T-x}(T) D(t,T;e)}}{\ExF{t}{D(t,T;e)}} \,.
\]%

The above definition may be simplified by a suitable choice of the measure under which we take the expectation. In particular, we can consider the collateralized $T$-forward measure $\mathbb{Q}^{T;e}$ defined by means of the following Radon-Nikodym derivative
\[
Z_t(T;e) := \left.\frac{d \mathbb{Q}^{T;e}}{d \mathbb{Q}}\right|_{{\cal F}_t} := \frac{\ExFT{t}{}{D(0,T;e)}}{P_0(T;e)} = \frac{D(0,t;e) P_t(T;e)}{P_0(T;e)}
\]%
which is a positive $\mathbb{Q}$-martingale, normalized so that $Z_0(T;e)=1$.

Thus, for any payoff $\phi_T$, perfectly collateralized at overnight rate $e_t$, we can express prices as expectations under the collateralized $T$-forward measure, and we get
\[
\ExFT{t}{}{\phi_T D(t,T;e)} = P_t(T;e) \,\ExFT{t}{T;e}{\phi_T} \,.
\]%
In particular, we can write collateralized forward Libor rates as
\begin{equation}
\label{eq:forward}
F_t(T,x;e) = \ExFT{t}{T;e}{L_{T-x}(T)} \,.
\end{equation}%

One-period forward rates $F_t(T,x;e)$, along with multi-period ones (swap rates), are actively traded on the market. Once collateralized zero-coupon bonds are derived, we can bootstrap forward rate curves from such quotes. See, for instance, \cite{Bianchetti2009} or \cite{PallaviciniTarenghi} for a discussion on bootstrapping algorithms.

Basis swaps are an interesting product that became more popular after the marked switched to a multicurve structure. In fact in a basis swap there are two floating legs, one pays a Libor rate with a certain tenor and the other pays the Libor rate with a shorter tenor plus a spread that makes the contract fair at inception. More precisely the discounted payoff of a basis
swap which legs pay respectively a Libor rate with tenors $x<y$ with maturity $T=nx=my$ is given by
\begin{eqnarray*}
  && \sum_{i=1}^n D(t,T-(n-i)x;e) \;x\,\left(L_{T-(n-i-1)x}(T-(n-i)x)+K\right) \\
- && \sum_{j=1}^m D(t,T-(m-j)y;e) \;y\,L_{T-(m-j-1)y}(T-(m-j)y) \,.
\end{eqnarray*}%
It is clear that a part from being traded per se, this instrument is naturally present in the banks portfolios as result of the netting of opposite swap positions with different tenors.

\subsection{Modeling Constraints}
\label{sec:constraints}

Our aim is to setup a multiple-curve dynamical model starting from collateralized zero-coupon bonds $P_t(T;e)$, and Libor forward rates $F_t(T,x;e)$. As we have seen we can bootstrap the initial curves for such quantities from directly observed quotes in the market. Now, we wish to propose a dynamics that preserves the martingale properties satisfied by such quantities. Thus, without loss of generality, we can define collateralized zero-coupon bonds under the $\mathbb{Q}$ measure as
\[
\frac{dP_t(T;e)}{P_t(T;e)} = e_t \,dt - \sigma^P_t(T;e)^* \,dW_t
\]%
and Libor forward rates under the $\mathbb{Q}^{T;e}$ measure as
\[
dF_t(T,x;e) = \sigma^F_t(T,x;e)^* \,dZ_t^{T;e}
\]%
where $W$s and $Z$s are correlated standard (column) vector\footnote{In the following we will consider $N$-dimensional vectors as  $N\times 1$ matrices. Moreover given a matrix $A$, we will indicate $A^*$ its transpose, and if $B$ is another conformable matrix we indicate $AB$ the usual matrix product.} Brownian motions with correlation matrix $\rho$, and the volatility vector processes $\sigma^P$ and $\sigma^F$ may depend on bonds and forward Libor rates themselves.

The following definition of $f_t(T;e)$ is not strictly necessary, and we could keep working with bonds $P_t(T;e)$, using their dynamics. However, as it is customary in interest rate theory to model rates rather than bonds, we may try to formulate quantities that are closer to the standard HJM framework. In this sense we can define instantaneous forward rates $f_t(T;e)$, by starting from (collateralized) zero-coupon bonds, as given by
\[
f_t(T;e) := -\frac{\partial}{\partial T} \log P_t(T;e)
\]%
We can derive instantaneous forward-rate dynamics by It\^o lemma, and we obtain the following dynamics under the $\mathbb{Q}^{T;e}$ measure
\[
df_t(T;e) = \sigma_t(T;e) \,dW_t^{T;e}
\;,\quad
\sigma_t(T;e) := \frac{\partial}{\partial T} \,\sigma^P_t(T;e)
\]%
where partial differentiation is meant to be applied component-wise. 

Hence, we can summarize our modeling assumptions in the following way.
\begin{enumerate}
\item Since linear products (OIS, IRS, basis swaps...) can be expressed in terms of simpler quantities, namely collateralized zero-coupon bonds $P_t(T;e)$ and Libor forward rates $F_t(T,x;e)$, we focus on their modeling. 
\item Initial term structures for collateralized products may be bootstrapped from market data. 
\item For volatility and dynamics, we can write rates dynamics by enforcing suitable no-arbitrage martingale properties, namely
\begin{eqnarray}
\label{eq:mm}
df_t(T;e) = \sigma_t(T;e)^* \,dW_t^{T;e}
\;,\quad
dF_t(T,x;e) = \sigma^F_t(T,x;e)^* \,dZ_t^{T;e}.
\end{eqnarray}
\end{enumerate}

As we explained in the introduction, this is where the multiple-curve picture finally shows up: we have a curve with Libor based forward rates $F_t(T,x;e)$, that are collateral adjusted expectation of Libor market rates $L_{T-x}(T)$ we take as primitive rates from the market, and we have instantaneous forward rates $f_t(T;e)$ that are OIS based rates. OIS rates $f_t(T;e)$ are driven by collateral fees, whereas Libor forward rates $F_t(T,x;e)$ are driven both by collateral rates and by the primitive Libor market rates.

Now, the framework for multiple curves is ready. In the following section we propose a specific dynamics within this framework.

\section{Interest-Rate Modeling}
\label{sec:modeling}

We can now specialize our modeling assumptions to define a model for interest-rate derivatives which is on one hand flexible enough to calibrate the quotes of the MM, and on the other hand robust. Our aim is to to use a HJM framework using a single family of Markov processes to describe all the term structures and interest rate curves we are interested in.

In the literature many authors proposed generalizations of interest-rate models to include multiple yield curves. In particular we cite the first temptatives of \cite{Kijima2009} and \cite{Fujii2010}, followed by the extensions of the Libor Market Models proposed by \cite{Mercurio2010,Mercurio2012}, short-rate models based on multiplicative basis as in \cite{Henrard2013}, the extensions of the HJM framework proposed by \cite{Moreni2010,Moreni2014}, \cite{Crepey2012d,Crepey2014c}, and \cite{Fontana2014}. A survey of the literature can be found in \cite{Henrard2014}.

In such works the problem is faced in a pragmatic way by considering each forward rate as a single asset without investigating the microscopical dynamics implied by liquidity and credit risks. However, the hypothesis of introducing different underlying assets may lead to over-parametrization issues that affect the calibration procedure. Indeed, the presence of swap and basis-swap  quotes on many different yield curves is not sufficient, as the market quotes swaption premia only on few yield curves. For instance, even if the Euro market quotes one-, three-, six- and twelve-month swap contracts, liquidly traded swaptions are only those indexed to the three-month (maturity one-year) and the six-month (maturities from two to thirty years) Euribor rates. Swaptions referring to other Euribor tenors or to Eonia are not actively quoted. A similar line of reasoning holds also for caps/floors and other interest-rate options.

In order to solve such problem \cite{Moreni2010} introduces a parsimonious model to describe a multi-curve setting by starting from a limited number of (Markov) processes, so to extend the logic of the HJM framework to describe with a unique family of Markov processes all the curves we are interested in.

\subsection{Multiple-Curve Collateralized HJM Framework}
\label{sec:multicurve}

Let us summarize the basic requirements our model must fulfill:
\begin{itemize}
\item[i)] existence of OIS rates, which we can describe in terms of instantaneous forward rates $f_t(T;e)$;
\item[ii)] existence of Libor rates assigned by the market, typical underlyings of traded derivatives, with associated forwards $F_t(T,x;e)$;
\item[iii)] no arbitrage dynamics of the $f_t(T;e)$ and the $F_t(T,x;e)$ (both being $(T,e)$-forward measure martingales); 
\item[iv)] possibility of writing both $f_t(T;e)$ and $F_t(T,x;e)$ as functions of a common family of Markov processes, so that we are able to build parsimonious yet flexible models.
\end{itemize}

We stress that our approach models only quantities whose initial conditions can be bootstrapped in a model independent way from market quotes, and it includes implicitly the margining procedure within valuation equations.

We reformulate the results of \cite{Moreni2010} by taking into account that interest-rate products are collateralized. Hence, we choose under $\mathbb{Q}^{T;e}$ measure, the following dynamics.
\begin{eqnarray}
\label{eq:mainSDE}
df_t(T;e) &=& \sigma_t(T)^* dW_t^{T;e} \\\nonumber
\frac{dF_t(T,x;e)}{k(T,x)+F_t(T,x;e)} &=& \Sigma_t(T,x)^* dW_t^{T;e}
\end{eqnarray}%
where we introduce the families of (stochastic $N$-dimensional) volatility processes $\sigma_t(T)$ and $\Sigma_t(T,x),$ the vector of $N$ independent $\mathbb{Q}^{T;e}$-Brownian motions $W_t^{T;e},$ and the set of deterministic shifts $k(T,x),$ such that 
\[
\lim_{x \downarrow 0} x k(T,x) = 1 \,.
\]%
This limit condition ensures that the model approaches a standard default and liquidity free HJM model when the tenor goes to zero. We bootstrap $f_0(T;e)$ and $F_0(T,x;e)$ from market quotes.

In order to satisfy the fourth of teh above requirements, getting a model with a reduced number of common driving factors in the spirit of HJM approaches, it is sufficient to conveniently tie together the volatility processes $\sigma_t(T)$ and $\Sigma_t(T,x)$ through a third process $\sigma_t(u,T,x)$.
\begin{equation}
\label{eq:commonVol}
\sigma_t(T) := \sigma_t(T;T,0)
\;,\quad
\Sigma_t(T,x) := \int_{T-x}^T\!\!\!\! \sigma_t(u;T,x)\,du \,.
\end{equation}

Under this parametrization the OIS curve dynamics is the very same as the risk-free curve in an ordinary HJM framework. Indeed, we have for linearly-compounding forward rates 
\[
\frac{dE_t(T,x;e)}{1/x+E_t(T,x;e)}= \int_{T-x}^T\sigma_t(u)^* \,du \;dW^{T;e}_t \,.
\]%
In the generalized version of the HJM framework proposed by \cite{Moreni2010} we have an explicit expression for both the collateralized zero-coupon bonds $P_t(T;e)$ and the Libor forward rates $F_t(T,x;e)$. The first result is a direct consequence of modeling the OIS curve as the risk-free curve in a standard HJM framework, while the second result can be achieved only if a particular form of the volatilities is selected. We obtain this if we generalize the approach of \cite{Ritchken1995} by introducing the following separability constraint
\begin{equation}
\label{eq:separableVol}
\sigma_t(u,T,x) := h(t)q(u,T,x)g(t,u) \,,
\end{equation}
\[
g(t,u) := \exp\left\{-\int_t^u \!\!a(s)ds\,\right\}
\;,\quad
q(u;u,0) := Id
\]%
where $h_t$ is a $N\times N$ matrix process, $q(u,T,x)$ is a deterministic $N\times N$ diagonal matrix function, and $a(s)$
is a deterministic $N$ dimensional vector function. The condition on $q(u;T,x)$ being the identity matrix, when $T=u$ ensures that a standard HJM framework holds for collateralized zero-coupon bonds.

We can work out an explicit expression for the Libor forward rates, by plugging the expression of the volatilities into equation~\eqref{eq:mainSDE}. We obtain 
\begin{multline}
\label{eq:separableLnF}
\log \left(\frac{k(T,x)+F_t(T,x;e)}{k(T,x)+F_0(T,x;e)}\right) = \\
G(t,T-x,T;T,x)^*\left(X_t+Y_t \left(G_0(t,t,T) - \frac{1}{2}\,G(t,T-x,T;T,x)\right)\right)
\end{multline}%
where the stochastic vector process $X_t$ and the auxiliary matrix process $Y_t$ are defined under the $\mathbb{Q}$ measure as in the ordinary HJM framework
\begin{align*}
X^i_t &= \sum_{k=1}^N \int_0^t g_i(s,t) \left(h_{ik,s} \,dW_{k,s} + (h_s^*h_s)_{ik}\int_s^t g_k(s,y) \,dy \;ds \right)
\;,\quad
i=1\dots N \\
Y_t^{ik} &= \int_0^t g_i(s,t) (h_s^*h_s)_{ik} g_k(s,t) \,ds
\;,\quad
i,k=1\dots N
\end{align*}%
and
\[
G_0(t,T_0,T_1) = \int_{T_0}^{T_1}g(t,s) \,ds
\;,\quad
G(t,T_0,T_1,T,x) = \int_{T_0}^{T_1}q(s,T,x)g(t,s) \,ds
\]%
Furthermore it can be shown that the processes $X^i$ and $Y^{ik}$ follow 
\begin{equation}\label{eq:xyHJM}
\begin{aligned}
dX^i =& \left(\sum_{k=1}^N Y^{ik}_t - a^i(t) X_t^i\right) \,dt + h_t^* \,dW_t \\
dY^{ik}_t =& \left( (h_t^*h_t)_{ik}-(a^i(t)+a^k(t))Y_t^{ik} \right) \,dt \,.
\end{aligned}
\end{equation}

It is worth noting that the integral representation of forward Libor volatilities given by equation~\eqref{eq:commonVol}, together with the common separability constraint given in equation~\eqref{eq:separableVol} are sufficient conditions to ensure the existence of a reconstruction formula for all OIS and Libor forward rates based on the very same family of Markov processes.Indeed, we can write the reconstruction formula for OIS rates as given by
\begin{multline}
\label{eq:separableLnE}
\log\left(\frac{\frac{1}{x}+E_t(T,x;e)}{\frac{1}{x}+E_0(T,x;e)}\right) =\\
G_0(t,T-x,T)^* \left(X_t+Y_t \left(G_0(t,t,T) - \frac{1}{2}\,G_0(t,T-x,T)\right)\right) \,.
\end{multline}%
Moreover we have the following formula for the OIS istantaneous forward rate
\begin{equation}
\label{eq:instaforward}
f_t(T) = f_0(T)+g(t,T)^*\left(X_t+Y_tG_0(t,t,T)\right) \,.
\end{equation}

In the next section we derive three different dynamics within the HJM collateralized framework, and we will use them in the numerical experiments of Section~\ref{sec:numerics}.

\subsection{Model Specifications}
\label{sec:specs}

We are interested in some specification of this model, in particular a variant of the Hull and White model, a variant of the Cheyette model and the Moreni and Pallavicini model.

The Hull and White model (hereafter HW) described in \cite{HullWhite} is the simplest one, and is obtained by choosing
\begin{equation}
\label{eq:parameters}
h(t)\doteq R
\;,\quad
q(u,T,x)\doteq Id
\;,\quad
a(s)\doteq a
\;,\quad
\kappa(T,x) \doteq \frac{1}{x}
\end{equation}
where $a$ is a constant vector, and $R$ is the Cholesky decomposition of the correlation matrix that we want our $X_t$ vector to have. In this case we obtain as $\sigma_t(u;T,x)$ process
\[
\sigma_t(u;T,x)= R \cdot e^{ - a (u-t) } 
\]
where the exponential is intended to be component-wise. Then we note that $X_t$ is a mean reverting Gaussian process while the $Y_t$ process is deterministic.

In order to model implied volatility smiles, we can add a stochastic volatility process to our model, as shown in \cite{Moreni2012}. In particular, we can obtain a variant of the Cheyette model (hereafter Ch) described in \cite{Cheyette1992} by considering a common square-root process for all the entries of $h$, as in \cite{Trolle2009}. More precisely we replace $h(t)$ in \eqref{eq:parameters} with
\[
h(t)\doteq\sqrt{v_t} R
\]
where $a$ and $R$ are defined as before. The volatility $v_t$ is a process with the following dynamics:
\begin{equation}
\label{eq:vola}
dv_t = \eta \left( 1 - v_t \right)\,dt + \nu_0 \left( 1 + (\nu_1-1)e^{-\nu_2 t} \right) \sqrt{v_t} \,dZ_t
\;,\quad
v_0 = {\bar v}
\end{equation}
where $Z_t$ is a Brownian motion correlated to $W_t$, so that we obtain as $\sigma_t(u;T,x)$ process
\[
\sigma_t(u;T,x) = \sqrt{v_t} R \cdot e^{ - a (u-t) } \,.
\]

As the last specification of the framework we consider the model \cite{Moreni2012} (hereafter MP) which adopts a different shift $k(T,x)$, and introduces a dependence on the tenor in the volatility process. 
\begin{equation}
h(t)\doteq\sqrt{v_t} R
\;,\quad
q(u,T,x)^{i,i}\doteq e^{x\eta^i} 
\;,\quad
a(s)\doteq a
\;,\quad
\kappa(T,x) \doteq \frac{e^{-\gamma x}}{x}
\end{equation}
where $a$ and $R$ are defined as before. The volatiltiy $v_t$ is defined by \eqref{eq:vola}. Here, we have for the $\sigma_t(u;T,x)$ process
\[
\sigma_t(u;T,x) =\sqrt{v_t} R \cdot e^{\eta x - a (u-t) }
\]%

To better apreciate the difference between the Ch and MP models it is useful to compute the quantity $\frac{P_t(t+x;e)}{P_t(t+x;x)}$, where $x$ is a tenor and 
\[
P_t(t+x;x):=\frac{1}{1+xF_t(t+x,x;e)}.
\]%
By means of equations~\eqref{eq:separableLnF} and~\eqref{eq:separableLnE}, we obtain that for the HW model and the Ch model the quantity  $\frac{P_t(t+x;e)}{P_t(t+x;x)}$ is deterministic and equals to
\[
\frac{P_t(t+x,e)}{P_t(t+x,x)}=\frac{1+xF_0(t+x,x)}
{1+xE_0(t+x,x)}
\]%
while for the MP model we have that
\begin{equation*}
\begin{aligned}
\frac{P_t(t+x,e)}{P_t(t+x,x)}
\end{aligned}
\end{equation*}
is a stochastic process. Hence, if we consider the following quantity, which represents the time normalized difference between two forward rate with different tenors,
\begin{equation}
\label{eq:beta}
\beta_t(x_1,x_2;e) := \frac{1}{x_2} \log\left(\frac{\frac{1}{x_2}+E_t(t+x_2,x_2;e)}{\frac{1}{x_2}+F_t(t+x_2,x_2;e)}\right) - \frac{1}{x_1} \log\left(\frac{\frac{1}{x_1}+E_t(t+x_1,x_1;e)}{\frac{1}{x_1}+F_t(t+x_1,x_1;e)}\right)
\end{equation}%
we have that in the HW model and in the Ch model $\beta_t(x_1,x_2;e)$ is deterministic while in the MP model is a stochastic quantity. This suggests that the MP model should be able to better capture the dynamics of the basis between two rates with different tenors.
  

\subsection{Numerical Results}
\label{sec:numerics}

We apply our framework to simple but relevant products: an IRS and a basis swap. We analyze the impact of the choice of an interest rate model on the portfolio valuation, in particular we measure:
\begin{enumerate}
\item the dependency of the price on the correlations between interest-rates and credit spreads, the so-called wrong-way risk;
\item the impact of the so called gap risk in an otherwise perfectly collateralized deal, due to the presence of a cure period of $\delta$ days.
\end{enumerate}

We model the market risks by simulating the following processes in a multiple curve HJM model under the pricing measure $\mathbb{Q}$. The overnight rate $e_t$ and the Libor forward rates $F_t(T;e)$ are simulated according to the dynamics given in Section~\ref{sec:multicurve}. Mantaining the same notation of the aforementioned section, we choose $N=2$, and for our numerical experiments we use a HW model, a Ch model and a MP model, all calibrated to swaption at-the-money volatilities listed on the Euro market. Table~\ref{tab:summary} summarizes the properties of the considered models. As we have already noted, the Ch model introduces a stochastic volatility and hence has an increased number of parameters with respect to the HW model. The MP model aims at better modeling the basis between rates with different tenors, while keeping the model parsimonious in terms of extra parameters with respect to the Ch model.

\begin{table}
\begin{center}
\begin{tabular}{|c|cc|ccc|}\hline
Model              & Factors & Pars & Rate   & Volatility & Basis      \\\hline\hline
HullWhite          & 2       & 5    & Stoch. & Constant   & Time Dep.  \\\hline
Cheyette           & 3       & 12   & Stoch. & Stoch.     & Time Dep.  \\\hline
MoreniPallavicini  & 3       & 14   & Stoch. & Stoch.     & Stoch.     \\\hline
\end{tabular}
\end{center}
\caption{\label{tab:summary}
Summary of the characteristics of the models used for the numerical experiments. In particular for each model is displayed the number of stochastic factors, the number of free parameters for the calibration, and the properties of rate, volatility and basis processes.}
\end{table}

\subsubsection{Calibration Results}
\label{sec:calibration_results}

\begin{figure}
\begin{center}\hspace*{-0.5cm}
\includegraphics[width=0.375\textwidth]{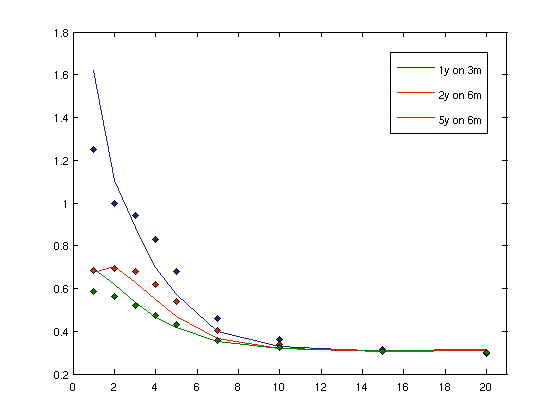}\hspace*{-0.5cm}
\includegraphics[width=0.375\textwidth]{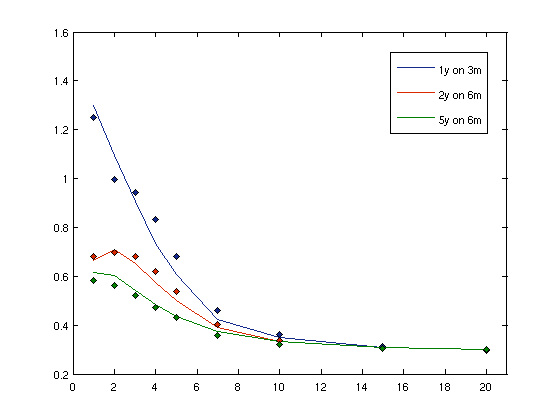}\hspace*{-0.5cm}
\includegraphics[width=0.375\textwidth]{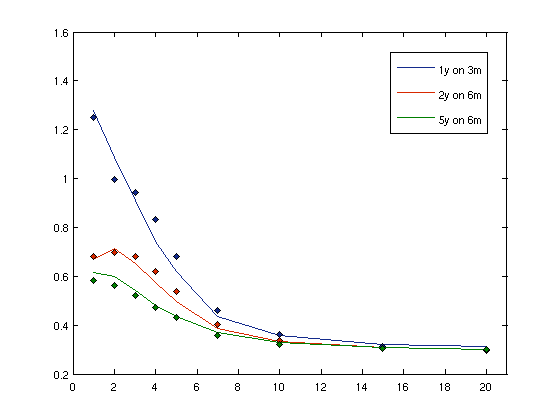}
\end{center}
\caption{\label{fig:atmvola}
At-the-money swaption volatilities. On the horizontal axis swaption expiries, on the vertical axis swaption volatilities. The curves correspond to swaptions with tenors of 1y (on 3m Euribor rate), and 2y, 5y (on 6m Euribor rate). The dots are market quotes. Left panel: HW model. Central panel: Ch model. Right panel: MP model.}
\end{figure}

\begin{figure}
\begin{center}\hspace*{-0.5cm}
\includegraphics[width=0.375\textwidth]{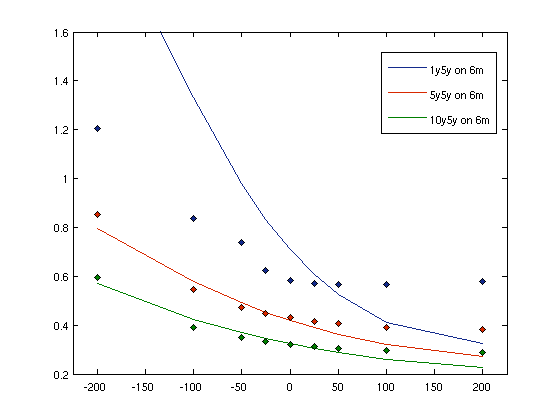}\hspace*{-0.5cm}
\includegraphics[width=0.375\textwidth]{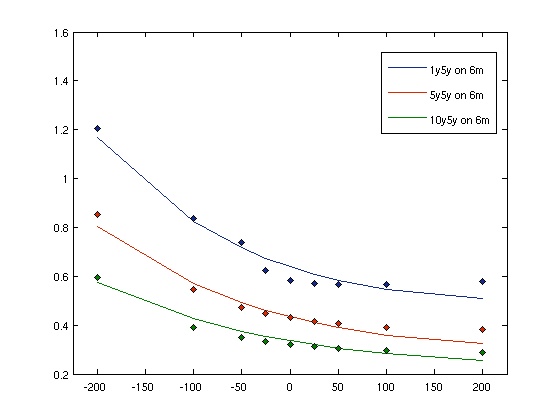}\hspace*{-0.5cm}
\includegraphics[width=0.375\textwidth]{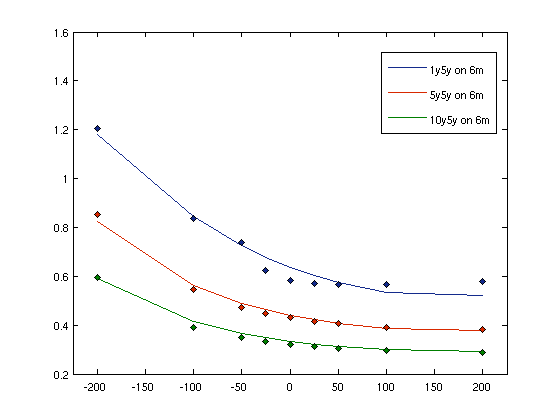}
\end{center}
\caption{\label{fig:atmsmile}
Swaption volatility smile. On the horizontal axis differences between swaption strike rate and underlying forward swap-rate, on the vertical axis swaption volatilities. The curves correspond to swaptions with tenor of 5y and expiries of 1y, 5y and 10y. The dots are market quotes. Left panel: HW model. Central panel: Ch model. Right panel: MP model.}
\end{figure}

The result of the calibration are shown in Figures~\ref{fig:atmvola} and~\ref{fig:atmsmile}. From these figures it is clear that the HW model is able to reproduce the at-the-money quotes but is not able to correctly reproduce the volatility smile. On the other hand the introduction of a stochastic volatility process helps recovering the market data smile and thus the Ch and the MP models have similar results in properly fitting the smile.

For what concerns the credit part, the default intensities of the investor and the counterparty are given by two CIR++ processes $\lambda^i_t$

\begin{equation*}
\begin{aligned}
&\lambda^i_t=y^i_t+\psi^i(t)\\
&dy^i_t=\zeta^i(\mu^i-y^i_t) \,dt + \nu^i \sqrt{y^i_t} \,dZ^{i,c}_t
\;,\quad
i\in\{I,C\}
\end{aligned}
\end{equation*}%
and they are calibrated to the market data shown in \cite{BrigoCapponiPallaviciniPapatheodorou}. In particular, two different market settings are used in the numerical examples: the medium risk and the high risk settings. The correlations among the risky factors are induced by correlating the Brownian motions as in \cite{BrigoPallavicini2007}.

\subsubsection{Wrong-Way Risk}
\label{sec:wwr}

The wrong-way risk (WWR) represent how the derivative price changes by varying the dependency between market and credit risks. In our case we calculate WWR for IRS and basis swaps. In the first case the dependency between market and credit risks is expressed in term of the correlation between the default intensities and the overnight rate, while in the second case we use the correlation between $\beta_t$, as given by Equation~\eqref{eq:beta}, and the default intensities.

We start analyzing the impact of wrong-way risk on the bilateral adjustment, namely CVA plus DVA, of IRS and basis swaps when collateralization is switched off, namely we want to evaluate Equation~\eqref{eq:masterF} when
\[
\alpha_t\doteq 0.
\]%

\begin{figure}
\begin{center}\hspace*{-0.5cm}
\includegraphics[width=0.55\textwidth]{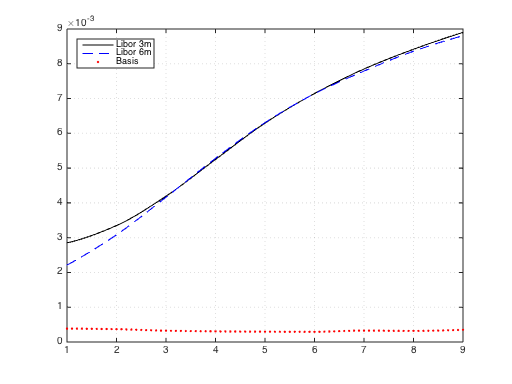}\hspace*{-0.5cm}
\includegraphics[width=0.55\textwidth]{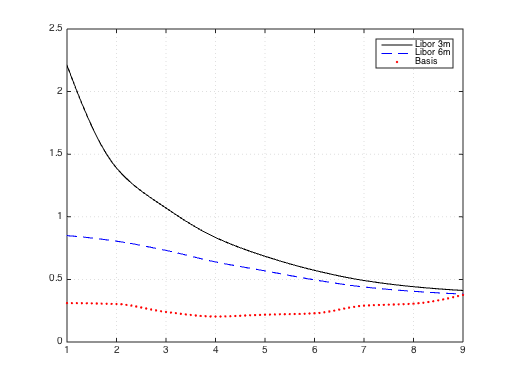}
\end{center}
\caption{\label{fig:volarates}
Volatility of 6m and 3m Libor rates and of $\beta_t(x_{\rm 3m},x_{\rm 6m};e)$ the proxy for the 6m/3m Libor basis. On the horizontal axis expiry $t$, on the vertical axis volatilities obtained via a Monte Carlo simulation. Left panel: normal volatilities. Right panel: log-normal volatilities.}
\end{figure}

\begin{figure}
\begin{center}
\includegraphics[width=0.55\textwidth]{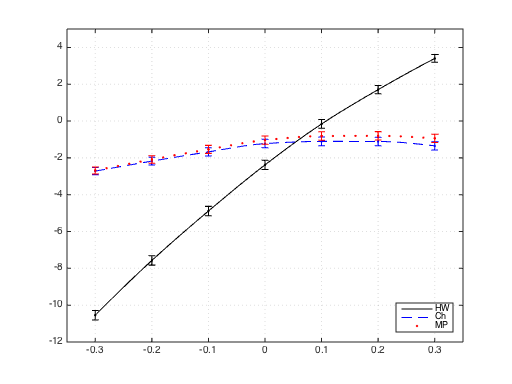}
\end{center}
\vspace*{-0.5cm}
\caption{\label{fig:wwrIRS}
Wrong-way risk for different models. On the horizontal axis correlation among credit and market risks; on the vertical axis the bilateral adjustment, namely CVA + DVA for a 10y IRS receiving a fix rate and paying 6m Libor. Values are in basis points.
}
\end{figure}

In Figure~\ref{fig:volarates} we show the volatilities of both Libor rates relative to the studied products and the basis $\beta_t$, 
while in Figure~\ref{fig:wwrIRS} we show the wrong way risk (WWR) as variation of the bilateral adjustment with respect to market-credit correlation for a ten years IRS receiving a fix rate yearly and paying 6m Libor twice a year.  We limit correlation to low values such as 0.3 also because it is unnatural to expect a higher correlation between a central market rate such as Libor or OIS and a specific counterparty credit spread. 
From our analysis, it is clear that for a product not subject to the basis dynamics we have that the big difference among the model is the presence of a stochastic volatility. In fact we can see that the Ch model and the MP model are almost indistinguishable while the results of the HW model are far from the stochastic volatility ones. Moreover we can observe that all the models have the same trend, i.e. the bilateral adjustment grows as correlation increase. In fact this can be explained by the fact that an higher correlation means that the deal will be more profitable when it will be more risky (since we are receiving the fixed rate and paying the floating one), hence the bilateral adjustment will be bigger. 

\begin{figure}
\begin{center}
\includegraphics[width=0.55\textwidth]{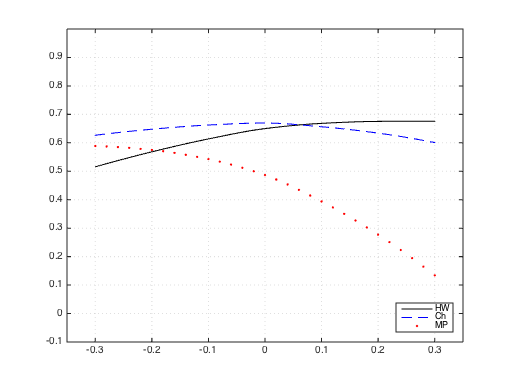}
\end{center}
\vspace*{-0.5cm}
\caption{
Wrong-way risk for different models. On the horizontal axis correlation among credit and market risks; on the vertical axis the bilateral adjustment, namely CVA + DVA for a 10y basis swap receiving 3m Libor plus spread  and paying 6m Libor. Values are in basis points. 
}
\label{fig:wwrBasis}
\end{figure}

\begin{figure}
\begin{center}
\includegraphics[width=0.55\textwidth]{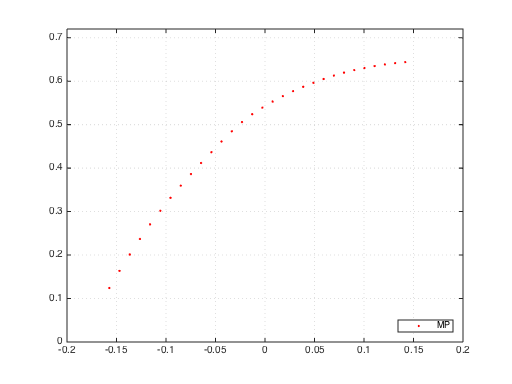}
\end{center}
\vspace*{-0.5cm}
\caption{
Wrong-way risk for MP model. On the horizontal axis correlation among default intensities and $\beta_t(x_{\rm 3m},x_{\rm 6m};e)$; on the vertical axis the bilateral adjustment, namely CVA + DVA for a 10y basis swap receiving 3m Libor plus spread  and paying 6m Libor. Values are in basis points. Right 
}
\label{fig:wwrMPbase}
\end{figure}

In Figure~\ref{fig:wwrBasis} we show instead the variation of the bilateral adjustment for a ten years basis swap  receiving 3m Libor plus spread and paying 6m Libor. In this case we see that as said before the HW model and the Ch model don't have a basis dynamic and hence the curve represented is almost flat. On the other hand the MP model is able to capture the dynamics of the basis and hence we can see that the more the basis is correlated with the credit risk  the smaller becomes the bilateral adjustment.
In Figure~\ref{fig:wwrMPbase} we focus on the MP model and we show how the wrong way risk depends on the correlation between the defaault intensities and $\beta_t(x_{\rm 3m},x_{\rm 6m};e)$.

\subsubsection{Gap Risk}
\label{sec:gr}

As a final analysis we studied the gap risk of IRS and basis swaps when collateralization is switched on, but initial margin is not exchanged, namely we use Equation~\ref{eq:masterFgap} with various values for $\alpha_t$. In particular we choose a cure period of $\delta=10$ days.

\begin{figure}
\begin{center}\hspace*{-0.5cm}
\includegraphics[width=0.375\textwidth]{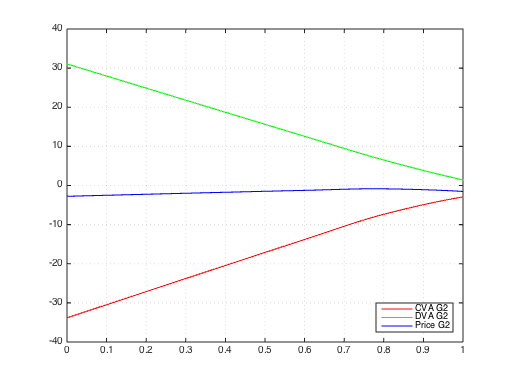}\hspace*{-0.5cm}
\includegraphics[width=0.375\textwidth]{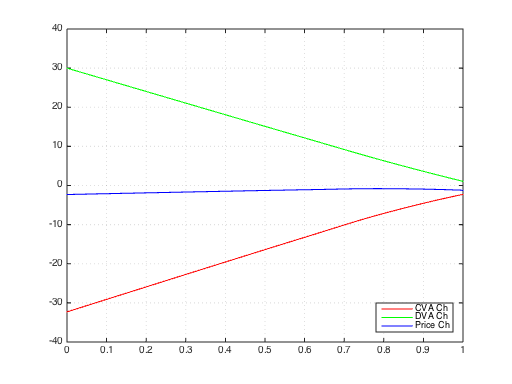}\hspace*{-0.5cm}
\includegraphics[width=0.375\textwidth]{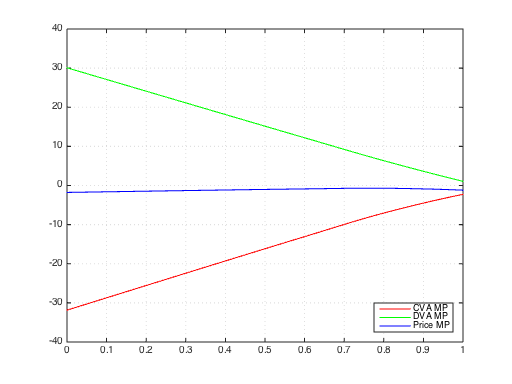}
\end{center}
\caption{\label{fig:alphaIRS}
CVA and DVA for an IRS in presence of cure period at various levels of collateralization. On the horizontal $\alpha_t$, i.e. the fraction of the close-out value covered by the collateral account, on the vertical axis the values of CVA, DVA and the price of the product all expressed in basis points. Left panel: HW model. Central panel: Ch model. Right panel: MP model.}
\end{figure}

\begin{figure}
\begin{center}\hspace*{-0.5cm}
\includegraphics[width=0.375\textwidth]{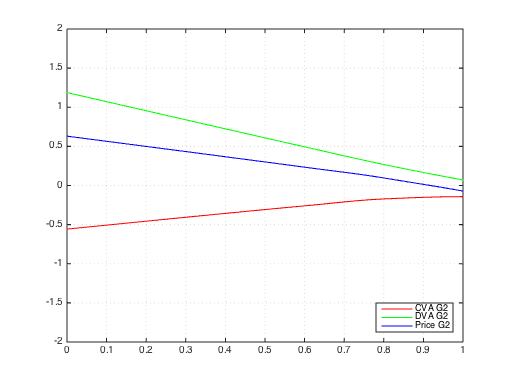}\hspace*{-0.5cm}
\includegraphics[width=0.375\textwidth]{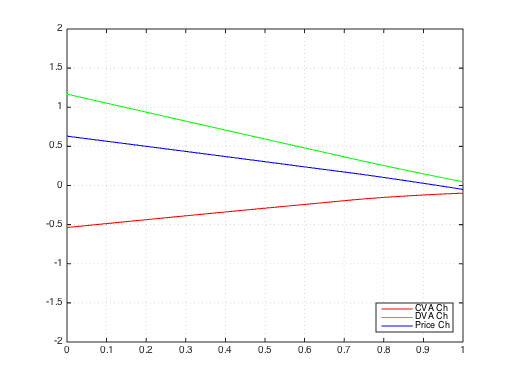}\hspace*{-0.5cm}
\includegraphics[width=0.375\textwidth]{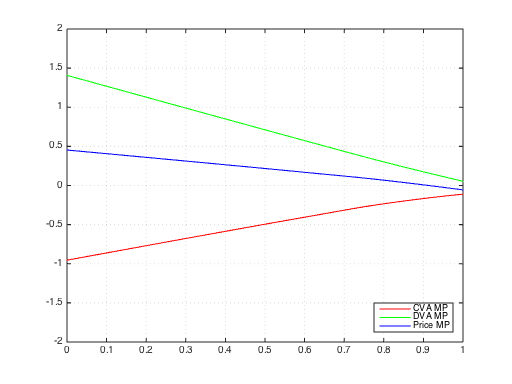}
\end{center}
\caption{\label{fig:alphabasis}
CVA and DVA for a basis swap in presence of cure period at various levels of collateralization. On the horizontal $\alpha_t$, i.e. the fraction of the close-out value covered by the collateral account, on the vertical axis the values of CVA, DVA and the price of the product all expressed in basis points. Left panel: HW model. Central panel: Ch model. Right panel: MP model.}
\end{figure}

First in Figures~\ref{fig:alphaIRS} and \ref{fig:alphabasis} we show how the collateralization procedure impacts the price of our products, i.e. how changes in $\alpha_t$ affect the CVA and DVA component of the price in presence of a delay. In particular we see that as expected high levels of collateralization reduce default risk adjustments, but even with perfect collateralization we have a residual riskdue to the cure period. 

\begin{figure}
\begin{center}\hspace*{-0.5cm}
\includegraphics[width=0.55\textwidth]{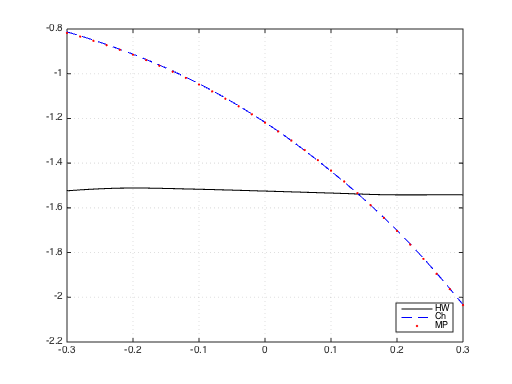}
\end{center}
\vspace*{-0.5cm}
\caption{
Residual bilateral adjustment due to gap risk. On the horizontal axis correlation among credit and market risks; on the vertical axis the bilateral adjustment, namely CVA + DVA for a 10y IRS receiving a fix rate and paying 6m Libor. Values are in basis points
}
\label{fig:gaprisk}
\end{figure}

Next we specialized our analysis to the case:
\[
\alpha_t\doteq 1
\]%
 The results for IRS are reported in Figure~\ref{fig:gaprisk}. In particular we see that in the case of an IRS the prices computed with the HW model are flat with respect to the correlation, and this is because while pricing with full collateralization in presence of margin period of risk what is left unhedged is the molatility of the product in the cure period. Since the HW model has a static volatility for the Libor rates, the volatility of the IRS payoff during the cure period is also static and hence shows little variation with respect to the canges in correlation among the market and credit risk factors. On the other hand models like Ch and MP, with a dynamic volatility structure, are susceptible to wrong-way risk and they show very similar results.   

\begin{figure}
\begin{center}\hspace*{-0.5cm}
\includegraphics[width=0.55\textwidth]{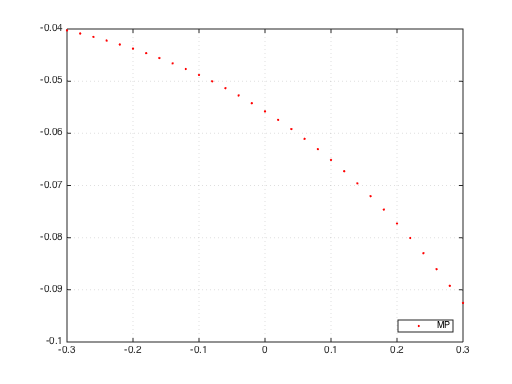}
\end{center}
\vspace*{-0.5cm}
\caption{
Residual bilateral adjustment due to gap risk. On the horizontal axis correlation among credit and market risks; on the vertical axis the bilateral adjustment, namely CVA + DVA for a 10y basis swap receiving 3m Libor plus spread  and paying 6m Libor. Values are in basis points
}
\label{fig:gapbasis}
\end{figure}

Lastly we analized the gap risk of a basis swap. In this case we see that the remaining risk is very small, one orders of magnitude less than the IRS, due to the fact that the basis between the rates involved has little volatility. In particular in this case the error of the numerical procedure used doesn't allow further investigation. On the other hand the gap risk is so small that is possible to neglect it and hence there is no need of implementing an initial margin procedure. In Figure~\ref{fig:gapbasis} we report the result for the MP model that showed some pattern beyond the numerical error. 

\section{Conclusions and Further Works}

In this paper we presented a detailed analysis of interest rate derivatives valuation under credit risk and collateral modeling. In particular, we used the framework in \cite{Perini2011} to define a pricing framework where we can specify different collateralization policies and we can include the possibility of a delay in the close-out procedure. Within this framework we reformulate the multiple-curve model of \cite{Moreni2010}. We discussed in a numerical section the impact of a stochastic basis in pricing credit valuation adjustments for IRS and basis swaps. 

This work is a starting point to understand which interest-rate products require a stochastic basis model. In particular, we could extend the analysis to compare different stochastic basis models.

In such a context funding costs enter the picture in a more comprehensive way. Some initial suggestions in this respect were given in \cite{BrigoPallavicini2013} and \cite{BrigoPallavicini2014}.

%

\newpage

\bibliographystyle{plainnat}
\bibliography{hjm_collateral}

\end{document}